\documentclass[amsmath,amssymb,prl,final,showpacs,twocolumn]{revtex4-1}
\usepackage{bm,hyphenat,xspace}
\usepackage{graphicx,epsfig}
\usepackage{color}

\begin{document}

\title {Comment on
"Is a Trineutron Resonance Lower in Energy than a Tetraneutron Resonance?"
}

\author{A.~Deltuva}
\email{arnoldas.deltuva@tfai.vu.lt}
\affiliation
{Institute of Theoretical Physics and Astronomy,
Vilnius University, Saul\.etekio al. 3, LT-10257 Vilnius, Lithuania
}
\author{R. Lazauskas}
\affiliation{IPHC, IN2P3-CNRS/Universit\'e Louis Pasteur BP 28,
F-67037 Strasbourg Cedex 2, France}

%\author{J. Carbonell}
%\affiliation{Institut de Physique Nucl\'eaire, Universit\'e Paris-Sud, IN2P3-CNRS, 91406 Orsay Cedex, France}

%\received{\today}
%\received{29 June, 2016}
%\pacs{21.30.-x, 21.45.-v}

 \maketitle

%\section{Introduction \label{sec:intro}}
%Study of few-neutron resonances presented in
%ef.~\cite{PhysRevLett.118.232501} provided results incompatible
%with rigorous few-body calculations. In this Comment we point out
%serious shortcomings in the framework of Ref.~\cite{PhysRevLett.118.232501}
%related to treatment of the few-body continuum.
%in particular the threshold effects.

%a \clearpage

The quantum Monte Carlo  study~\cite{PhysRevLett.118.232501}
of few-neutron resonant states provided results incompatible
with rigorous few-body calculations
\cite{lazauskas:4n,PhysRevC.93.044004,deltuva:18b}.
In this Comment we point out
serious shortcomings in the framework of Ref.~\cite{PhysRevLett.118.232501},
leading to  misinterpretation of unbound few-body systems.

Study of unbound few-neutron systems
\cite{PhysRevLett.118.232501} followed quite popular
strategy consisting of two steps: (i) make the system bound
 with additional attractive potential, controlled by
strength parameter $V_0$;
(ii) extrapolate the resulting binding energy
 to the physical limit in continuum at $V_0=0$.
Two different ways for step (i) have been employed in
Ref.~\cite{PhysRevLett.118.232501}: 1) adding an external
trap potential and fixing center-of-mass (c.m) of the
system; 2) enhancing the $nn$ interaction by factor
$\alpha=1+V_0$. Such procedure is sound if (a) the
calculated bound state is physical and it  evolves into
resonance, and (b) the analytic continuation to different
Reemann sheet with resonance is performed correctly, taking into
account threshold effects.

 We argue that {\em both} these conditions are not satisfied in
Ref.~\cite{PhysRevLett.118.232501}. For definiteness we consider
the four-neutron ($4n$) system.
Additional attraction may generate a
 bound dineutron with energy $E_d<0$,  which then defines
the stability threshold for tetraneutrons: only those with
$E_{4n}\le\,E_d$ in the trap  (or those with $E_{4n}\le\,2E_d$ for
the enhanced force) are stable. Otherwise, even in the case $E_{4n} < 0$
they can decay into
dineutron plus two infinitesimally slow neutrons moving around the
common mass center (trap)
 or into two dineutrons (produced by enhanced force).

Our study
reveals that a bound dineutron emerges in trap with radius
$R_{\mathrm{WS}}=6$~fm
 and potential depth $V_0\approx-0.09$~MeV only,
or when the enhancement factor $\alpha$ in the ${}^1S_0$ wave
 exceeds $\approx\,1.1$
(these values  slightly depend on the underlying $nn$ potential).
However, $4n$ states declared to be bound tetraneutrons with
$E_{4n}\to\,0$
in Ref.~\cite{PhysRevLett.118.232501} were found only at
significantly larger absolute values of $V_0\approx\,-1.2$ MeV and
$\alpha\approx\,1.3$. For such Hamiltonians the dineutrons are
already well bound, thus, the lowest-energy state of the system is
not true (stable) bound state, but continuum state that
asymptotically looks like dineutron in a trap plus two slow
peripheral neutrons. It appears that
Ref.~\cite{PhysRevLett.118.232501}  ignored this effect 
in the presumed $E_{4n}\approx\,0$ region, which is decisive for the
extrapolation. The tetraneutron states of Ref.~\cite{PhysRevLett.118.232501}
are above the stability threshold and therefore are not true bound
states but most probably represent some discretized continuum states that do not
evolve into a resonance. Extrapolation of their energies does not
lead to proper resonance energy.

Furthermore, a caution is  needed in extrapolation procedure itself
if real bound states are calculated, since trajectory of a bound
state evolving into continuum state involves branching at each
threshold with discontinuity in the second derivative of energy
with respect to strength parameter~\cite{Kukulin}.  Polynomial
extrapolations~\cite{PhysRevLett.118.232501} neglect this
discontinuity  and therefore are conceptually incorrect.

\begin{figure}[!]
\begin{center}
\includegraphics[scale=0.55]{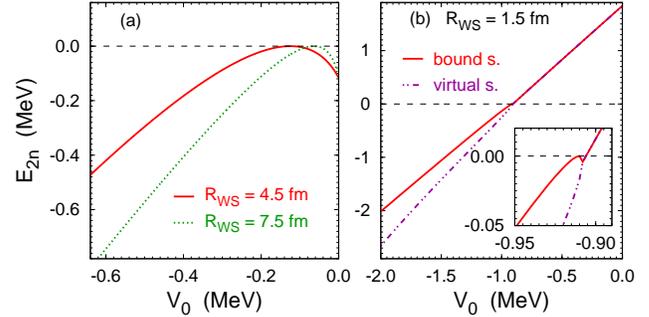}
\end{center}
\caption{\label{fig:1}
${}^1S_0$ dineutron pole trajectories in Wood-Saxon
traps with  given range parameters
for realistic (a) and two-Gaussian (b) potentials.}
\end{figure}

We show two examples in Fig.~\ref{fig:1}
corresponding to the ${}^1S_0$ virtual state
for a realistic potential and to the resonance of the two-Gaussian
potential \cite{PhysRevLett.118.232501}. Obtained ${}^1S_0$ pole
trajectories  have a typical bending shape, resulting  $-0.12$ MeV
virtual state energy, in sharp contrast with the positive 0.1 MeV
value of  Ref.~\cite{PhysRevLett.118.232501}. The latter is
obtained by a polynomial extrapolation neglecting the
near-threshold bending region. The resonance of the two-Gaussian
potential does not necessarily evolve from the ground state in the
trap. In  favorable case a linear extrapolation, avoiding the
input from the near-threshold region, may give a reasonable
estimation for the energy of a narrow resonance. However, presence
of  a branching point at the threshold, as shown in Fig.~\ref{fig:1}~(inset),
 produces highly nonlinear effects rendering
naive extrapolation procedures mathematically unjustified.

\begin{acknowledgments}
Authors acknowledge discussions with J. Carbonell.
A.D. acknowledges the support  by the Alexander von Humboldt Foundation
under Grant No. LTU-1185721-HFST-E.
\end{acknowledgments}

%\clearpage
%%%%%%%%%%%%%%%%%%%%%%%%%%%%%%%%%%%%%%%%%%%%%%%%%%%%%%%%%%%%%%%%%%%%%%%%%%%%%
%\bibliographystyle{prsty}
%\bibliography{abbrev,pre80,80-89,90-99,200x,4N,ad,book,atomic,lowk,n34}

\end{document}